\documentstyle[aps,pra,preprint,eqsecnum]{revtex}

\def\be{\begin{eqnarray}}
\def\ee{\end{eqnarray}}
\def\l{\langle}

\def\A{{\cal A}}

\newcommand{\QED}{\hspace*{\fill}\mbox{\rule[0pt]{1.2ex}{1.2ex}}}

\begin{document}

\tightenlines

\title{Universal state inversion and concurrence in arbitrary dimensions}

\author{
Pranaw Rungta,$^{(1)}$ V.~Bu\v{z}ek,$^{(2)}$\cite{padd}
Carlton M.~Caves,$^{(1)}$\\
M.~Hillery,$^{(3)}$ and G.~J.~Milburn$^{(4)}$}

\address{
$^{(1)}$Center for Advanced Studies, Department of Physics and Astronomy,\\
University of New Mexico, Albuquerque, NM 87131--1156, USA\\
$^{(2)}$The Erwin Schr\"{o}dinger Institute for Mathematical Physics,\\
Boltzmanngasse 9, A-1090 Wien, Austria\\
$^{(3)}$Department of Physics and Astronomy, Hunter College of CUNY,\\
695 Park Avenue, New York, NY 10021, USA\\
$^{(4)}$Centre for Quantum Computer Technology,\\
The University of Queensland, QLD 4072, Australia
}
\date{2001 June~2}
\maketitle

\begin{abstract}
Wootters [Phys.\ Rev.\ Lett.\ {\bf 80}, 2245 (1998)] has given an
explicit formula for the entanglement of formation of two qubits
in terms of what he calls the {\it concurrence\/} of the joint
density operator.  Wootters's concurrence is defined with the help
of the superoperator that flips the spin of a qubit.  We
generalize the spin-flip superoperator to a ``universal
inverter,'' which acts on quantum systems of arbitrary dimension,
and we introduce the corresponding generalized concurrence for
joint pure states of $D_1\times D_2$ bipartite quantum systems. We
call this generalized concurrence the {\it I-concurrence} to
emphasize its relation to the universal inverter.  The universal
inverter, which is a positive, but not completely positive
superoperator, is closely related to the completely positive
universal-NOT superoperator, the quantum analogue of a classical
NOT gate.  We present a physical realization of the universal-NOT
superoperator.
\end{abstract}

\section{Introduction}
\label{sec:intro}

Entanglement plays a central role in quantum information
theory~\cite{Lo1998,Nielsen2000}.  Perhaps the most important measure
of entanglement for bipartite systems is the entanglement of
formation \cite{Bennett1996,Cbennett1996}.  For a bipartite pure
state $|\Psi^{AB}\rangle$, the entanglement of formation is given by
the entropy of the marginal density operators, $\rho_A$ and $\rho_B$,
of systems $A$ and $B$.  For a bipartite mixed state $\rho_{AB}$, the
entanglement of formation is given by the minimum average marginal
entropy of ensemble decompositions of $\rho_{AB}$.

Hill and Wootters \cite{Hill1997} introduced another measure of
entanglement, called the {\it concurrence}, for pairs of qubits.
The concurrence is defined with the help of a superoperator
${\cal S}_2$, whose action on a qubit density operator
$\rho={1\over2}(I+\vec P\cdot\vec\sigma)$ is to flip the spin
of the qubit:
\begin{equation}
{\cal S}_2(\rho)
=\sigma_y\rho^\ast\sigma_y
={1\over2}(I-\vec P\cdot\vec\sigma)\;.
\label{eq:spinflip2}
\end{equation}
Here $\rho^\ast$ is the complex conjugate (or transpose) of $\rho$
relative to the eigenbasis of $\sigma_z$.  The concurrence of a
pure state $|\Psi_{AB}\rangle$ of two qubits is defined to be
\cite{Hill1997}
\begin{equation}
C_2(\Psi_{AB})\equiv\sqrt{\Bigl\langle\Psi_{AB}\Bigl|{\cal
S}_2\otimes{\cal S}_2
(|\Psi_{AB}\rangle\langle\Psi_{AB}|)\Bigr|\Psi_{AB}\Bigr\rangle}
=|\langle\Psi_{AB}|\sigma_y\otimes\sigma_y|\Psi_{AB}^\ast\rangle|\;.
\label{eq:firstC2}
\end{equation}
Wootters \cite{Wootters1998} showed that the entanglement of
formation of an arbitrary two-qubit mixed state $\rho_{AB}$ can be
written in terms of the minimum average pure-state concurrence, where
the minimum is taken over all ensemble decompositions of $\rho_{AB}$,
and he derived an explicit expression for this minimum average
pure-state concurrence.  Wootters called this minimum the concurrence
of the mixed state.  Uhlmann \cite{Uhlmann2000} introduced a
generalization of Wootters's concurrence to higher dimensions, which
we discuss further in Sec.~\ref{subsec:uinverter}.

In this paper we generalize the notion of concurrence to pairs of
quantum systems of arbitrary dimension, in a way different from
Uhlmann's.  We show in Sec.~\ref{sec:uinverter} that if the
concurrence is to be generated by a product superoperator, as in the
expression~(\ref{eq:firstC2}), then the only suitable superoperator
to go into the tensor product is what we call the ``universal
inverter.''  For a $D$-dimensional quantum system, which we call a
``qudit,'' we denote the universal inverter by ${\cal S}_D$.  The
action of the universal inverter on a qudit state $\rho$ is given by
\begin{equation}
{\cal S}_D(\rho)=\nu_D(I-\rho)\;,
\end{equation}
where $\nu_D$ is a positive constant.  Acting on a pure qudit state
$|\psi\rangle$, the universal inverter maps $|\psi\rangle$ to a
multiple of the maximally mixed state in the subspace orthogonal to
$|\psi\rangle$.  The universal inverter has been used previously in
studies of the separability of mixed states by Horodecki and
Horodecki \cite{Horodecki1999}.

The corresponding generalized concurrence for a joint pure state
$|\Psi_{AB}\rangle$ of a $D_1\times D_2$ system, in analogy to
Eq.~(\ref{eq:firstC2}) for qubits, is given by
\begin{equation}
C(\Psi_{AB})\equiv
\sqrt{\Bigl\langle\Psi_{AB}\Bigl|{\cal S}_{D_1}\otimes{\cal S}_{D_2}
(|\Psi_{AB}\rangle\langle\Psi_{AB}|)\Bigr|\Psi_{AB}\Bigr\rangle}
=2\nu_{D_1}\nu_{D_2}
[1-{\rm tr}(\rho_A^2)]\;.
\label{eq:firstC}
\end{equation}
Thus, for pure states, this generalized concurrence is simply
related to the purity of the marginal density operators.  A
sensible choice for the constant $\nu_D$, consistent with the
concurrence for qubits, is $\nu_D=1$.  We call the generalized
concurrence~(\ref{eq:firstC}) the {\it I-concurrence\/} to
emphasize its relation to the universal inverter and also to
distinguish it from a generalized concurrence introduced by
Uhlmann \cite{Uhlmann2000}.

The universal inverter is a natural generalization to higher
dimensions of the qubit spin flip.  Only for $D=2$, the spin flip,
does the universal inverter map pure states to pure states.  The
universal inverter cannot be realized as a quantum dynamics, because
though it is a positive superoperator, it is not completely positive.
In Sec.~\ref{subsec:tpops} we explore a one-parameter family of
trace-preserving superoperators that are closely related to the
universal inverter, and we show that the completely positive member
of this family that is closest to the universal inverter is the
universal-NOT superoperator \cite{Buzek1999,Buzek2000}.  The
universal-NOT is thus a physically realizable quantum analogue of the
classical NOT gate. The action of the universal-NOT, denoted ${\cal
G}_{\rm NOT}$, on a qudit state is given by
\begin{equation}
{\cal G}_{\rm NOT}(\rho)={1\over D^2-1}(DI-\rho)\equiv\rho^{\rm NOT}\;.
\end{equation}
In Sec.~\ref{sec:uNOT} we give two physical realizations of the
universal-NOT, one in terms of the quantum information distributor
introduced by Braunstein, Bu\v{z}ek, and Hillery
\cite{Braunstein2000} and the other in terms of a measurement of the
isotropic POVM followed by state inversion.

The paper concludes in Sec.~\ref{sec:conclusion} with a brief
discussion that includes the natural extension of I-concurrence to
mixed states.

\section{Universal Inverter}
\label{sec:uinverter}

In this section we first review, in Sec.~\ref{subsec:spinflip},
Wootters's spin-flip operation for a qubit and how it leads to an
entanglement measure called the concurrence for an arbitrary pure
state of two qubits~\cite{Wootters1998}.  The main result of this
paper is to generalize the spin flip to a superoperator that we
call the {\it universal inverter\/}.  The universal inverter is
defined in all Hilbert-space dimensions, and it leads to a
generalized concurrence for joint pure states of two quantum
systems of arbitrary dimension.  In Sec.~\ref{subsec:uinverter} we
formulate the requirements for the universal inverter and explore
some of its properties, in Sec.~\ref{subsec:proof} we show that
these requirements pick out a unique universal inverter up to a
constant multiple, and in Sec.~\ref{subsec:tpops} we consider
trace-preserving superoperators that are closely related to the
universal inverter.

The formalism we use for superoperators has been used extensively
in open-systems theory \cite{Alicki1987}.  The particular notation
we use can be found in Ref.~\cite{Caves1999} and is summarized
briefly in Appendix~\ref{app:supop}, along with a description of
several superoperators that play key roles in our discussion.  In
contrast to Ref.~\cite{Caves1999}, we use $\odot$, instead of
$\otimes$, to denote the slot into which one inserts the operator
on which a superoperator acts, reserving $\otimes$ to denote
tensor products between quantum systems.  This superoperator
formalism has been used to analyze entanglement in
Ref.~\cite{Rungta2000}.

We refer to the two subsystems of a bipartite system as systems
$A$ and $B$.  Where necessary for clarity, we use subscripts $A$,
$B$, and $AB$ to distinguish quantities belonging to the
subsystems and to the joint system. To reduce notational clutter,
however, we omit these subscripts on pure states, denoting pure
states of a single system by a lower-case Greek letter, e.g.,
$|\psi\rangle$, and joint pure states of a bipartite system by an
upper-case Greek letter, e.g., $|\Psi\rangle$.

\subsection{Spin flip and qubit concurrence}
\label{subsec:spinflip}

A spin flip for a single qubit is effected by the anti-unitary
operator $\sigma_y{\cal C}=-{\cal C}\sigma_y$, where ${\cal C}$
denotes complex conjugation in the eigenbasis of $\sigma_z$. Acting
on a state vector $|\psi\rangle$ or an operator $A$, the anti-unitary
complex conjugation operator gives ${\cal
C}|\psi\rangle=|\psi^*\rangle$ or ${\cal C}A={A^\ast}{\cal C}$, where
$|\psi^*\rangle$ and $A^*$ denote complex conjugation of the state or
operator in the eigenbasis of $\sigma_z$.  An anti-unitary operator
that satisfies $\Theta^2=\pm1$, i.e., $\Theta^\dagger=\pm\Theta$, is
called a {\it conjugation\/}; conjugations are ordinarily introduced
in quantum mechanics to represent time reversal.  Complex conjugation
in some orthornormal basis is a conjugation because ${\cal C}^2=1$,
and spin flip is a conjugation because $(\sigma_y{\cal
C})^\dagger={\cal C}^\dagger\sigma_y^\dagger={\cal
C}\sigma_y=-\sigma_y{\cal C}$. For a description of other properties
and uses of anti-linear operators, see Ref.~\cite{Messiah1968}.

Promoted to an operator on operators, the spin flip becomes an {\it
anti-linear\/} superoperator $\sigma_y{\cal C}\odot{\cal C}\sigma_y$,
which acts on operators according to $\sigma_y{\cal C}A{\cal
C}\sigma_y=\sigma_y A^\ast\sigma_y$.  Since we are only interested in
the operation of the spin flip on Hermitian operators, where complex
conjugation is equivalent to transposition, we can replace this
anti-linear superoperator with the corresponding linear superoperator
\begin{equation}
\label{eq:spinflip1}
{\cal S}_2=\sigma_y\odot\sigma_y\circ{\cal T}_2\;,
\end{equation}
where ${\cal T}_2$ denotes transposition in the eigenbasis of
$\sigma_z$ (see Appendix~\ref{app:supop}).  The subscript~2
distinguishes the spin flip and transposition in two dimensions
from the similar quantities for arbitrary dimensions that we
introduce later in this section.

The action of the spin-flip superoperator on an arbitrary qubit
density operator, $\rho={1\over2}(I+\vec P\cdot\vec\sigma)$, is to
invert the Bloch vector $\vec P$ through the origin, as in
Eq.~(\ref{eq:spinflip2}).  Since inversion commutes with
rotations, representing unitary operators, we have immediately
that ${\cal S}_2$ commutes with all unitary operators $U$, i.e.,
${\cal S}_2\circ U\odot U^\dagger=U\odot U^\dagger\circ{\cal
S}_2$.

For a quantum state $\rho$ of a two-qubit system, the spin-flipped
density operator, distinguished by a tilde, is
\begin{equation}
\label{eq:spinflip}
\tilde{\rho}={\cal S}_{2}\otimes{\cal S}_{2}(\rho)
=\sigma_y\otimes\sigma_y\rho^\ast\sigma_y\otimes\sigma_y\;.
\end{equation}
Hill and Wootters \cite{Hill1997} defined the concurrence of a
two-qubit pure state, $\rho=|\Psi\rangle\langle\Psi|$, to be
\begin{equation}
C_2(\Psi)\equiv\sqrt{{\rm tr}(\rho\tilde\rho)}
=\sqrt{\Bigl\langle\Psi\Bigl|{\cal S}_2\otimes{\cal S}_2
(|\Psi\rangle\langle\Psi|)\Bigr|\Psi\Bigr\rangle}
=|\langle\Psi|\sigma_y\otimes\sigma_y|\Psi^\ast\rangle|\;.
\label{eq:C2}
\end{equation}
The joint pure state can be written in terms of a Schmidt
decomposition,
\begin{equation}
|\Psi\rangle=
a_1|e_1\rangle\otimes|f_1\rangle+a_2|e_2\rangle\otimes|f_2\rangle\;,
\end{equation}
where $|e_j\rangle$ and $|f_j\rangle$ are the orthonormal
eigenvectors of the marginal density operators for the two qubits
and $a_1$ and $a_2$ are the (positive) square roots of the
corresponding eigenvalues.  Since ${\cal S}_2$ commutes with all
unitary operators, the concurrence $C_2(\Psi)$ is unchanged by
local unitary transformations.  This means that $C_2(\Psi)$ is a
function only of $a_1$ and $a_2$; it is easy to verify that
$C_2(\Psi)=2a_1a_2$.  As noted by Wootters, the concurrence can
serve as a measurement of entanglement: it is invariant under
local unitary transformations, as any good measure of entanglement
should be, and it varies smoothly from $0$ for pure product states
to $1$ for maximally entangled pure states.

Wootters \cite{Wootters1998} went on to show that the concurrence
can also be used to measure the entanglement of mixed states of
two qubits.  He showed that the entanglement of formation of an
arbitrary two-qubit mixed state $\rho$ can be written in terms of
the minimum average pure-state concurrence of ensemble
decompositions of $\rho$, and he derived an explicit expression
for this minimum in terms of the eigenvalues of $\rho\tilde\rho$.
Wootters called the minimum the concurrence of the mixed state.

Uhlmann \cite{Uhlmann2000} based his generalization of concurrence on
the fact that the spin flip is a conjugation, defining concurrence in
arbitrary dimensions in terms of a conjugation $\Theta$.  Following
Uhlmann, we call his concurrence the {\it $\Theta$-concurrence}, and
we discuss it further in the next subsection.

\subsection{Universal inverter and I-concurrence}
\label{subsec:uinverter}

Our goal in this paper is to generalize the spin-flip
superoperator ${\cal S}_2$  for a qubit to a superoperator ${\cal
S}_D$ that acts on qudit states and generates a generalized
concurrence for $D_1\times D_2$ bipartite quantum systems.  The
spin-flip superoperator has several important properties that we
might wish its generalization to retain:
\begin{enumerate}
\item ${\cal S}_2$ maps Hermitian operators to Hermitian operators.
\item ${\cal S}_2$ commutes with all unitary operators.
\item $\langle\Psi|{\cal S}_2\otimes{\cal S}_2
(|\Psi\rangle\langle\Psi|)|\Psi\rangle$ is nonnegative for all
joint pure states $|\Psi\rangle$ and goes to zero if and only if
$|\Psi\rangle$ is a product state.
\item ${\cal S}_2$ is a positive superoperator; i.e., it maps
positive operators to positive operators.
\item ${\cal S}_2$ is trace preserving.
\item ${\cal S}_2$ maps any pure state $|\psi\rangle\langle\psi|$
to the orthogonal pure state
$|\psi^\perp\rangle\langle\psi^\perp|$.
\item ${\cal S}_2$ is derived from a conjugation $\Theta$, i.e., an
antiunitary operator satisfying $\Theta^2=\pm1$.
\end{enumerate}
Property~1 guarantees that ${\cal S}_2\otimes{\cal S}_2$ maps
Hermitian operators to Hermitian operators (see
Appendix~\ref{app:HtoH}) and thus that the quantity
$\langle\Psi|{\cal S}_2\otimes{\cal S}_2
(|\Psi\rangle\langle\Psi|)|\Psi\rangle$ of property~3 is real.
Property~2 ensures that $C_2(\Psi)$ is unchanged by local unitary
transformations, as an entanglement measure should be. Property~3
makes $C_2(\Psi)$ well defined, by ensuring that the quantity
inside the square root is nonnegative, and it sets the zero so
that pure product states, but no other pure states, have vanishing
concurrence.

In generalizing the spin flip to higher dimensions, we want the
generalized concurrence of a pure state
$\rho=|\Psi\rangle\langle\Psi|$ of a $D_1\times D_2$ bipartite
system to be defined as for qubits, i.e.,
\begin{equation}
C(\Psi)\equiv
\sqrt{\Bigl\langle\Psi\Bigl|{\cal S}_{D_1}\otimes{\cal S}_{D_2}
(|\Psi\rangle\langle\Psi|)\Bigr|\Psi\Bigr\rangle}\;.
\label{eq:ditcon}
\end{equation}
It is clear that the analogues of properties 1--3 are desirable
properties of ${\cal S}_D$, for the same reasons as for qubits,
and it turns out that they are sufficient to pick out a unique
superoperator ${\cal S}_D$ up to a constant multiple.

The upshot of this discussion is that we require ${\cal S}_D$
to have the following properties:
\begin{itemize}
\item[$1'$.]${\cal S}_D$ maps Hermitian operators to Hermitian
operators.
\item[$2'$.] ${\cal S}_D$ commutes with all unitary operators.
\item[$3'$.] $\langle\Psi|{\cal S}_{D_1}\otimes{\cal S}_{D_2}
(|\Psi\rangle\langle\Psi|)|\Psi\rangle$ is nonnegative for all
joint pure states $|\Psi\rangle$ and goes to zero if and only if
$|\Psi\rangle$ is a product state.
\end{itemize}
The only superoperator that has these three properties is
\begin{equation}
{\cal S}_D=\nu_D({\bf I}-{\cal I})\;,
\label{eq:uinverter}
\end{equation}
where ${\bf I}$ is the unit superoperator relative to the
left-right action, ${\cal I}$ is the unit superoperator relative
to the ordinary action, and $\nu_D$ is an arbitrary real constant.
For the considerations in Sec.~\ref{subsec:tpops}, we allow
$\nu_D$ to have a dependence on $D$.  For purposes of defining a
generalized concurrence, however, $\nu_D$ should be independent of
$D$; otherwise the generalized concurrence of  joint pure state
could be changed simply by adding extra, unused dimensions to one
or both systems.

We show that ${\cal S}_D$ is the {\it only\/} superoperator
allowed by properties $1'$--$3'$ in Sec.~\ref{subsec:proof}.  For
the remainder of this subsection, we show that ${\cal S}_D$ does
satisfy properties $1'$--$3'$, and we spell out some of its other
properties and properties of the corresponding concurrence.
Notice first that ${\cal S}_D$ takes an operator $A$ to
\begin{equation}
{\cal S}_D(A)=\nu_D[\,{\bf I}(A)-{\cal I}(A)]=
\nu_D[{\rm tr}(A)I-A]\;,
\end{equation}
from which it is clear that ${\cal S}_D$ satisfies properties
$1'$ and $2'$.  If $A$ is a density operator $\rho$, we get
\begin{equation}
{\cal S}_D(\rho)=\nu_D(I-\rho)\;.
\end{equation}
Since $I-\rho$ is a positive operator for any $\rho$, we have
immediately that ${\cal S}_D$ is a positive superoperator provided
that $\nu_D$ is positive.  The generalized concurrence is
indifferent to a change in the sign of $\nu_D$, so we are free to
choose $\nu_D$ to be positive, which we do henceforth, thus making
${\cal S}_D$ positive.  If $\nu_D=1/(D-1)$, ${\cal S}_D$ is trace
preserving; this trace-preserving normalization is useful for the
considerations of Sec.~\ref{subsec:tpops}, but we see below that
$\nu_D=1$ is a more reasonable normalization to use for the
generalized concurrence $C(\Psi)$. Finally, ${\cal S}_D$ maps a
pure state $\rho=|\psi\rangle\langle\psi|$ to a positive multiple
of the projector orthogonal to $\rho$:
\begin{equation}
{\cal S}_D(|\psi\rangle\langle\psi|)=
\nu_D(I-|\psi\rangle\langle\psi|)\;.
\end{equation}
It is this property that prompts us to call ${\cal S}_D$ the {\it
universal inverter}.  We call the corresponding generalized
concurrence~(\ref{eq:ditcon}) the {\it I-concurrence\/} to emphasize
its connection with inversion.  Other properties of ${\cal S}_D$,
which follow directly from the corresponding properties of ${\bf I}$
and ${\cal I}$ (see Appendix~\ref{app:supop}), are that ${\cal S}_D$
is Hermitian relative to the ordinary action, i.e., ${\cal
S}_D^\times={\cal S}_D$, and that it changes sign under sharping,
i.e., ${\cal S}_D^\#=-{\cal S}_D$.  The universal inverter has been
used previously by Horodecki and Horodecki \cite{Horodecki1999} to
provide a criterion for the separability of mixed states.

We now see that properties 4--6 of the qubit spin flip survive, in
amended form, in its generalization:
\begin{itemize}
\item[$4'$.] ${\cal S}_D$ is a positive superoperator.
\item[$5'$.] ${\cal S}_D$ is a positive multiple of a trace-preserving
superoperator, i.e., ${\cal S}_D^\times(I)=\nu_D(D-1)I$.
\item[$6'$.] ${\cal S}_D$ maps any pure state $|\psi\rangle\langle\psi|$
to a positive multiple of the projector onto the subspace
orthogonal to $|\psi\rangle$.
\end{itemize}
It is worth pointing out that if we added to properties $1'$--$3'$
the additional requirement that ${\cal S}_D$ map each pure state
to a multiple of some orthogonal state, then the superoperator of
Eq.~(\ref{eq:uinverter}) would trivially be the only possibility
for the universal inverter.

We still have to deal with property~$3'$.  For that purpose we need
the tensor-product superoperator
\begin{equation}
{\cal S}_{D_1}\otimes{\cal S}_{D_2}=
\nu_{D_1}\nu_{D_2}
({\bf I}\otimes{\bf I}-{\cal I}\otimes{\bf I}
-{\bf I}\otimes{\cal I}+{\cal I}\otimes{\cal I})\;.
\end{equation}
Applied to an arbitrary joint density operator $\rho_{AB}$, this
tensor-product superoperator gives
\begin{equation}
{\cal S}_{D_1}\otimes{\cal S}_{D_2}(\rho_{AB})=
\nu_{D_1}\nu_{D_2}
(I\otimes I-\rho_A\otimes I-I\otimes\rho_B+\rho_{AB})\;.
\end{equation}
Projecting back onto $\rho_{AB}$ gives
\begin{equation}
{\rm tr}\Bigl
(\rho_{AB}{\cal S}_{D_1}\otimes{\cal S}_{D_2}(\rho_{AB})\Bigr)=
\nu_{D_1}\nu_{D_2}
[1-{\rm tr}(\rho_A^2)-{\rm tr}(\rho_B^2)+{\rm tr}(\rho_{AB}^2)]\ge0\;.
\label{eq:mixedcon}
\end{equation}
The inequality here, which shows that the quantity in
property~$3'$ is nonnegative, is proved in Appendix~\ref{app:rho},
where it is also shown that the inequality is saturated if and
only if $\rho_{AB}=\rho_A\otimes\rho_B$ is a product state, with
$\rho_A$ or $\rho_B$ a pure state.  For a joint pure state
$\rho_{AB}$, this establishes property~$3'$.

It is useful to specialize Eq.~(\ref{eq:mixedcon}) to a joint pure
state $|\Psi\rangle$, in which case it becomes the square of the
pure-state I-concurrence:
\begin{equation}
C^2(\Psi)=
\Bigl\langle\Psi\Bigl|{\cal S}_{D_1}\otimes{\cal S}_{D_2}
(|\Psi\rangle\langle\Psi|)\Bigr|\Psi\Bigr\rangle
=2\nu_{D_1}\nu_{D_2}
[1-{\rm tr}(\rho_A^2)]\;.
\end{equation}
Thus the I-concurrence measures the entanglement of a pure state in
terms of the purity, ${\rm tr}(\rho_A^2)={\rm tr}(\rho_B^2)$, of the
marginal density operators.  A joint pure state has a Schmidt
decomposition,
\begin{equation}
|\Psi\rangle=
\sum_j a_j|e_j\rangle\otimes|f_j\rangle\;,\quad a_j>0,
\end{equation}
in terms of which the squared I-concurrence becomes
\begin{equation}
C^2(\Psi)
=2\nu_{D_1}\nu_{D_2}\Biggl(1-\sum_j a_j^4\Biggr)
=4\nu_{D_1}\nu_{D_2}\sum_{j<k}a_j^2a_k^2\;.
\end{equation}
For defining a concurrence, one should choose the scaling factor
$\nu_D$ to be independent of $D$---otherwise, as noted above, the
pure state I-concurrence could be changed simply by adding extra,
unused dimensions to one of the subsystems---and to be consistent
with the qubit concurrence, one should choose $\nu_D=1$.  With
this choice the pure-state I-concurrence runs from zero for product
states to $\sqrt{2(M-1)/M}$, where $M=\min(D_1,D_2)$, for a
maximally entangled state.

Of the seven properties of the spin flip listed above, the first six
survive, some in amended form, in the universal inverter.  The
seventh, that ${\cal S}_2$ is derived from a conjugation, is not a
property of ${\cal S}_D$, because except in two dimensions, a
conjugation cannot commute with all unitaries and thus cannot serve
as the basis for a measure of entanglement.  Uhlmann's work on
conjugations \cite{Uhlmann2000} is valuable in that it generalizes to
all conjugations the expression that Wootters gives for the minimum
average pure-state concurrence of a bipartite density operator $\rho$
in terms of the eigenvalues of $\rho\tilde\rho$. Our results show,
however, that Uhlmann's $\Theta$-concurrence \cite{Uhlmann2000},
founded as it is on the use of conjugations, cannot serve as the
basis for a general measure of entanglement.

There is another interesting form of the universal inverter, which
makes a direct connection to the form~(\ref{eq:spinflip}) of the
spin flip.  Choosing an orthonormal basis $|e_j\rangle$, let
${\cal T}$ be the superoperator that transposes matrix
representations in this basis, and let ${\cal P}_A$ be the
superoperator projector, relative to the left-right action, which
projects onto the subspace of operators that are antisymmetric in
this basis.  We show in Appendix~\ref{app:supop} that
\begin{equation}
{\cal S}_D/\nu_D=2\,{\cal P}_A\circ{\cal T}\;.
\end{equation}
This form of the universal inverter has been given previously by
Horodecki and Horodecki \cite{Horodecki1999}.  For qubits, if we use
the eigenstates of $\sigma_z$ as the chosen basis, then the
antisymmetric operator subspace is spanned by the normalized operator
$\sigma_y/\sqrt2$, so the projector onto this subspace is ${\cal
P}_A=|\sigma_y)(\sigma_y|/2=\sigma_y\odot\sigma_y/2$. Thus in the two
dimensions the universal inverter becomes ${\cal
S}_2=\nu_2\sigma_y\odot\sigma_y\circ{\cal T}_2$, which agrees with
the spin flip if $\nu_2=1$.

\subsection{Derivation of universal inverter}
\label{subsec:proof}

We now show that the only superoperator that satisfies
properties~$1'$--$3'$ of the preceding subsection is the universal
inverter~(\ref{eq:uinverter}).  As we proceed through the proof,
we use ${\cal G}_D$ to denote the operator under consideration.

As we show in Appendix~\ref{app:HtoH}, property~$1'$ implies that
${\cal G}_D$ is left-right Hermitian, i.e., ${\cal G}_D={\cal
G}_D^\dagger$, and thus has an eigendecomposition
\begin{equation}
{\cal G}_D=
\sum_{\alpha}\mu_\alpha|\tau_\alpha)(\tau_\alpha|
=\sum_{\alpha}\mu_\alpha\tau_\alpha\odot\tau_\alpha^\dagger\;,
\label{eq:Geigendecomp}
\end{equation}
where the $\mu_\alpha$ are real (left-right) eigenvalues and the
operators $\tau_\alpha$ are the corresponding orthonormal
eigenoperators.

Property~$2'$ implies that
\begin{equation}
{\cal G}_D=U^\dagger\odot U\circ{\cal G}_D\circ U\odot U^\dagger
=\sum_{\alpha}\mu_\alpha
U^\dagger\tau_\alpha U\odot U^\dagger\tau_\alpha^\dagger U\;,
\end{equation}
which means that $U^\dagger\tau_\alpha U$ is an eigenoperator of
${\cal G}_D$, with eigenvalue $\mu_\alpha$, for any unitary operator
$U$.  This result can be restated as saying that the degenerate
eigensubspaces of ${\cal G}_D$ are invariant under all unitary
transformations.  We show in Appendix~\ref{app:eigensubspaces}
that the only operator subspaces that are invariant under all
unitary transformations are the one-dimensional subspace spanned
by the unit operator and the ($D^2-1$)-dimensional subspace of
tracefree operators.  As a consequence, ${\cal G}_D$ must have the
form
\begin{equation}
{\cal G}_D=\mu_D{\cal I}/D+\nu_D{\cal F}\;.
\end{equation}
Here ${\cal I}=I\odot I$ is the unit superoperator relative to the
ordinary action, ${\cal F}$ is the superoperator that projects onto
the subspace of tracefree operators when acting to the right
(see Appendix~\ref{app:supop}), $\mu_D$ is the eigenvalue of ${\cal
G}_D$ corresponding to the normalized eigenoperator $I/\sqrt D$, and
$\nu_D$ is the eigenvalue corresponding to all of the tracefree
operators.  Notice that ${\cal G}_D$ is Hermitian relative to the
ordinary action, i.e., ${\cal G}_D={\cal G}_D^\times$.

If we add $I/\sqrt D$ to a complete, orthonormal set of tracefree
operators, we obtain a complete, orthonormal set of operators,
so the unit superoperator in the left-right sense is given by
\begin{equation}
\label{eq:unit}
{\bf I}={\cal I}/D+{\cal F}\;,
\end{equation}
from which we get
\begin{equation}
{\cal G}_D=\eta_D{\cal I}+\nu_D{\bf I}\;,
\label{eq:GDform}
\end{equation}
where
\begin{equation}
\eta_D=(\mu_D-\nu_D)/D\;.
\end{equation}

Now we impose property~$3'$.  In doing so, it is sufficient to
consider the requirements of property~$3'$ in the case where the
two subsystems have the the same dimension $D$.  In this case the
tensor-product superoperator takes the form
\begin{equation}
{\cal G}_D\otimes{\cal G}_D=
\eta_D^2{\cal I}\otimes{\cal I}+
\eta_D\nu_D({\cal I}\otimes{\bf I}+{\bf I}\otimes{\cal I})
+\nu_D^2{\bf I}\otimes{\bf I}\;.
\end{equation}
Applying this superoperator to a joint density operator
$\rho_{AB}$ gives
\begin{equation}
{\cal G}_D\otimes{\cal G}_D(\rho_{AB})=
\eta_D^2\rho_{AB}+
\eta_D\nu_D(\rho_A\otimes I+I\otimes\rho_B)
+\nu_D^2I\otimes I\;,
\end{equation}
and projecting this back onto $\rho_{AB}$ yields
\begin{equation}
{\rm tr}\Bigl(\rho_{AB}{\cal G}_D\otimes{\cal G}_D(\rho_{AB})\Bigr)=
\eta_D^2{\rm tr}(\rho_{AB}^2)+
\eta_D\nu_D[{\rm tr}(\rho_A^2)+{\rm tr}(\rho_B^2)]
+\nu_D^2\;.
\end{equation}

Specializing to a joint pure state $|\Psi\rangle$, we get
\begin{equation}
\Bigl\langle\Psi\Bigl|{\cal G}_D\otimes{\cal G}_D
(|\Psi\rangle\langle\Psi|)\Bigr|\Psi\Bigr\rangle
=\eta_D^2+\nu_D^2+2\eta_D\nu_D{\rm tr}(\rho_A^2)
=(\eta_D\mp\nu_D)^2\pm2\eta_D\nu_D[1\pm{\rm tr}(\rho_A^2)]\;.
\label{eq:key}
\end{equation}
If $\eta_D\nu_D\ge0$, the top sign in Eq.~(\ref{eq:key}) shows
that the quantity in property~$3'$ is strictly positive, unless
$\eta_D=\nu_D=0$, a case of no interest.  If $\eta_D\nu_D<0$, the
bottom sign in Eq.~(\ref{eq:key}) shows that the quantity is
nonnegative and goes to zero if and only if $\eta_D=-\nu_D$ and
$\rho_A$ is pure, i.e., the joint pure state is a product state.
Thus it turns out that the quantity in property~$3'$ is
nonnegative for all superoperators of the form~(\ref{eq:GDform}),
but the only way to set the zero properly is to choose
$\eta_D=-\nu_D$, thus giving the universal inverter of
Eq.~(\ref{eq:uinverter}). The left-right eigenvalues of the
universal inverter are $\nu_D$ and
$\mu_D=D\eta_D+\nu_D=-(D-1)\nu_D$.

\subsection{Trace-preserving superoperators}
\label{subsec:tpops}

All superoperators of the form~(\ref{eq:GDform}) are proportional
to a trace-preserving superoperator, since
\begin{equation}
{\cal G}_D^{\times}(I)={\cal G}_D(I)=(\eta_D+D\nu_D)I\;.
\end{equation}
Requiring ${\cal G}_D$ to be trace preserving gives the condition
\begin{equation}
\eta_D=1-D\nu_D
\end{equation}
[$\mu_D=D-\nu_D(D^2-1)$], which allows us to eliminate one
parameter and to write the trace-preserving version of ${\cal
G}_D$ as
\begin{equation}
{\cal G}_{DT}=(1-D\nu_D){\cal I}+\nu_D{\bf I}\;.
\label{eq:Gtp}
\end{equation}
Acting on an arbitrary input state $\rho$, this superoperator
gives
\begin{equation}
{\cal G}_{DT}(\rho)=(1-D\nu_D)\rho+\nu_D I\;.
\label{eq:Gtprho}
\end{equation}
It is instructive to investigate this one-parameter family of
trace-preserving operators.

We first ask which of the trace-preserving
operators~(\ref{eq:Gtp}) are completely positive.  The condition
that a superoperator be completely positive is that its left-right
eigenvalues be nonnegative (see Appendix~\ref{app:supop}).  Thus
the condition for the complete positivity of ${\cal G}_{DT}$ is
that $\mu_D\ge0$ and $\nu_D\ge 0$, which is equivalent to
\begin{equation}
0\le\nu_D\le{D\over D^2-1}\;.
\end{equation}
When $\nu_D=0$, ${\cal G}_{DT}={\cal I}$ is the unit superoperator,
and when $\nu_D=D/(D^2-1)$,
\begin{equation}
\label{eq:gnot}
{\cal G}_{DT}={D\over D^2-1}{\cal F}
={1\over D^2-1}(D{\bf I}-{\cal I})
\equiv{\cal G}_{\rm NOT}
\end{equation}
is the universal-NOT superoperator \cite{Buzek1999,Buzek2000}. Notice
that the universal-NOT is a multiple of ${\cal F}$, the superoperator
whose right action projects onto the subspace of tracefree operators.
Since the dynamics of a quantum system must be completely positive,
the universal-NOT is the closest physical approximation to the
universal inverter in the one-parameter family~(\ref{eq:Gtprho}); it
is the quantum analogue of the classical NOT gate.  We present a
realization of the universal-NOT in Sec.~\ref{sec:uNOT}.

Another interesting completely positive superoperator occurs for
$\nu_D=1/(D+1)$:
\begin{equation}
{\cal G}_{DT}=
{1\over D+1}({\bf I}+{\cal I})
={1\over D}{\cal I}+{1\over D+1}{\cal F}
\equiv{\cal G}_{\rm AV}\;.
\label{eq:GAV}
\end{equation}
This superoperator was used to generate operator expansions in
Ref.~\cite{Rungta2000}, where it was shown that it is the unique
trace-preserving superoperator that satisfies ${\cal G}={\cal
G}^\dagger={\cal G}^\times={\cal G}^\#$ and commutes with all
unitaries.  In contrast, the universal inverter is the unique
superoperator that satisfies ${\cal G}={\cal G}^\dagger={\cal
G}^\times=-{\cal G}^\#$ and commutes with all unitaries.

As shown in Ref.~\cite{Rungta2000}, the superoperator ${\cal
G}_{\rm AV}$ is the trace-preserving version of the superoperator
that describes projection onto a random pure state,
\begin{equation}
{\cal G}_{\rm AV}
=D\int{d{\cal V}\over{\cal V}}\,
|\psi\rangle\langle\psi|\odot|\psi\rangle\langle\psi|\;,
\label{eq:GAV2}
\end{equation}
where $d{\cal V}$ is the unitarily invariant integration measure
on projective Hilbert space and ${\cal V}$ is the corresponding
total volume.  Projection onto a random pure state is the
measurement that results in the optimal estimation of the state of
the qudit \cite{Derka98}.  This estimated state is given by
the density operator
\begin{equation}
{\cal G}_{\rm AV}(\rho)=\frac{1}{D+1}(I+\rho)\;.
\end{equation}
The superoperator ${\cal G}_{\rm AV}$ returns in Sec.~\ref{sec:uNOT}
as an ingredient in one of the physical realizations of the
universal-NOT.

We now consider which of the trace-preserving
operators~(\ref{eq:Gtp}) are positive.  Letting $p_j$ be the
eigenvalues of the input density operator $\rho$, one sees that
the eigenvalues of ${\cal G}_{DT}(\rho)$ [Eq.~\ref{eq:Gtprho})]
are $(1-D\nu_D)p_j+\nu_D$. The condition that ${\cal G}_{DT}$ be
positive is that these eigenvalues be nonnegative for all input
eigenvalues $p_j$, which is equivalent to
\begin{equation}
0\le\nu_D\le{1\over D-1}\;.
\end{equation}
When $\nu_D=1/(D-1)$, ${\cal G}_{DT}$ becomes the trace-preserving
version of the universal inverter,
\begin{equation}
{\cal S}_{DT}={1\over D-1}({\bf I}-{\cal I})\;.
\label{eq:tpuinverter}
\end{equation}
The positive superoperators are convex combinations of
${\cal I}$ and ${\cal S}_{DT}$:
\begin{equation}
{\cal G}_{DT}=[1-\nu_D(D-1)]{\cal I}+\nu_D(D-1){\cal S}_{DT}\;.
\end{equation}
Notice that the universal-NOT can be written as
\begin{equation}
{\cal G}_{\rm NOT}={1\over2}({\cal S}_{DT}+{\cal G}_{\rm AV})\;.
\end{equation}

\section{Physical realizations of the universal-NOT}
\label{sec:uNOT}

In this section we give two physical realization of the universal-NOT
superoperator ${\cal G}_{\rm NOT}$ of Eq.~(\ref{eq:gnot}), the first
in terms of the quantum information distributor introduced by
Braunstein, Bu\v{z}ek, and Hillery \cite{Braunstein2000} and the
second in terms of a measurement of the isotropic POVM followed by
state inversion.

For the first, consider a qudit in a pure state
$\rho=|\psi\rangle\langle\psi|$. As shown in
Sec.~\ref{sec:uinverter}, the ideal inversion of this state is given
by
\begin{equation}
\label{eq:rperp}
{\cal S}_{DT}(\rho)=
{1\over D-1}({I}-\rho)\equiv
{\rho}^\perp
\;,
\end{equation}
where ${\cal S}_{DT}$ is the trace preserving version of the
universal inverter [see Eq.~(\ref{eq:tpuinverter})].  The inverted
state $\rho^\perp$ is the maximally mixed state in the
$(D-1)$-dimensional subspace orthogonal to the input state
$\rho=|\psi\rangle\l\psi|$.  Notice that by construction, ${\rm
tr}(\rho\rho^\perp)=0$ for pure input states.

As shown in Sec.~\ref{subsec:tpops}, the trace-preserving
universal inverter ${\cal S}_{\rm DT}$ is a positive, but not
completely positive superoperator and as such cannot be realized
physically.   In the one-parameter family of trace-preserving
inverters considered in Sec.~\ref{subsec:tpops}, the universal-NOT
superoperator ${\cal G}_{\rm NOT}$ of Eq.~(\ref{eq:gnot}) is the
closest completely positive superoperator to the universal
inverter.  We denote the physically possible inversion of the
state $\rho$ obtained using the universal-NOT as
\begin{equation}
\label{eq:inver}
\rho^{\rm NOT}\equiv{\cal G}_{\rm NOT}(\rho)={1\over D^2-1}(DI-\rho)\;.
\end{equation}

In order to realize the universal-NOT, we couple the qudit to be
inverted, denoted by $A$, to the quantum information distributor
(QID) introduced in Ref.~\cite{Braunstein2000}.  The QID is composed
of two ancilla qudits, $B$ and $C$, each of which has the same
dimension $D$ as qudit $A$.  To describe the universal inverter,
we introduce several operators and states for qudits.

First we need the conjugate ``position'' and ``momentum'' operators,
$x$ and $p$.  The eigenvectors of $x$ are denoted by $|x_k\rangle$,
\begin{equation}
\label{eq:xk}
x |x_{k}\rangle = x_k|x_{k}\rangle\;,
\end{equation}
with the eigenvalues given by $x_k=k\sqrt{2\pi/D}$; analogously, the
eigenstates of $p$ are denoted by $|p_k\rangle$,
\begin{equation}
\label{eq:pk}
p |p_{k}\rangle = p_k|p_{k}\rangle\;,
\end{equation}
with the eigenvalues given by $p_k=k\sqrt{2\pi/D}$.  We use units
such that the two operators are dimensionless.  The two sets of
eigenvectors, $\{|x_k\rangle\}$ and $\{|p_k\rangle\}$, form bases
in the qudit Hilbert space and are related by a discrete Fourier
transform,
\begin{eqnarray}
\label{eq:ft}
|x_k\rangle
&=&{1\over{\sqrt D}}\sum_{l=0}^{D-1}e^{-2\pi ikl/D}|p_l\rangle\;, \\
|p_l\rangle
&=&{1\over{\sqrt D}}\sum_{k=0}^{D-1}e^{2\pi ikl/D}|x_k\rangle \;.
\end{eqnarray}
The translation (shift) operators, defined by
\begin{equation}
R_{x}(n)=e^{-i x_np}\;,\qquad
R_{p}(m)=e^{i p_mx}\;,
\end{equation}
cyclically permute the basis vectors according to
\begin{eqnarray}
R_{x}(n)|x_{k}\rangle &=& |x_{(k+n){\rm mod}\,D}\rangle\;, \\
R_{p}(m)|p_{l}\rangle &=& |p_{(l+m){\rm mod}\,D}\rangle\;,
\end{eqnarray}
where the sums of indices are taken modulo $D$.

An orthonormal basis of $D^2$ two-qudit maximally entangled states
$|\Xi_{mn}\rangle$ is given by
\begin{equation}
\label{xix}
|\Xi_{mn}\rangle =
{1\over{\sqrt D}}
\sum_{k=0}^{D-1}
e^{2\pi imk/D} |x_{k}\rangle\otimes|x_{(k+n){\rm mod}\,D}\rangle\;,
\end{equation}
where $m,n=0,\dots,D-1$.  Using Eq.~(\ref{eq:ft}), we can rewrite
the states $|\Xi_{mn}\rangle$ in the joint momentum basis:
\begin{equation}
\label{xip}
|\Xi_{mn}\rangle=
{1\over{\sqrt D}}
\sum_{l=0}^{D-1}
e^{-2\pi inl/D}|p_{(m-l){\rm mod}\,D}\rangle\otimes|p_l\rangle\;.
\end{equation}
The state $|\Xi_{00}\rangle$ can be written as
\begin{equation}
\label{eq:Xi00}
|\Xi_{00}\rangle
=\frac{1}{\sqrt{D}}
\sum_{k=0}^{D-1}|x_{k}\rangle\otimes|x_{k}\rangle
=\frac{1}{\sqrt{D}}
\sum_{l=0}^{D-1}
|p_{-l\,{\rm mod}\,D}\rangle\otimes|p_{l}\rangle\;.
\end{equation}
It is interesting to note that the whole set of $D^2$ maximally
entangled states $|\Xi_{mn}\rangle$ can be generated from
$|\Xi_{00}\rangle$ by the action of {\em local\/} unitary operations
(shifts):
\begin{equation}
|\Xi_{mn}\rangle =
{R}_p(m)\otimes {R}_x(n) |\Xi_{00}\rangle \,.
\end{equation}

Now we are ready to describe the QID. \ The ancilla qudits, $B$
and $C$, are initially prepared in the state
\begin{equation}
\label{eq:phi}
|\Phi\rangle_{BC}=
\xi_1|\Xi_{00}\rangle_{BC}+\xi_2 |x_0\rangle_B\otimes|p_0\rangle_C\;.
\end{equation}
The phase freedom in $|\Phi\rangle_{BC}$ can be used to make
$\xi_1$ real and nonnegative, but then $\xi_2$ is in general
complex.  We do not use the freedom to make $\xi_1$ nonnegative,
thereby retaining for use below the ability to multiply both
$\xi_1$ and $\xi_2$ by $-1$.

Normalization of $|\Phi_{BC}\rangle$ imposes the constraint
\begin{equation}
1=\xi_1^2+|\xi_2|^2+{\xi_1(\xi_2+\xi_2^*)\over D}
=\xi_1^2+a^2+b^2+{2a\xi_1\over D}\;,
\label{eq:cons}
\end{equation}
where $\xi_2=a+ib$.  Solving for $\xi_1$, we get
\begin{equation}
\label{x1}
\xi_1=-{a\over D}+\sqrt{1-b^2-a^2{D^2-1\over D^2}}\;.
\end{equation}
We discard the other solution of the quadratic equation, because
it can be converted to this solution by multiplying both $\xi_1$
and $\xi_2$ by $-1$.  Since $\xi_1$ is real, we must have
\begin{equation}
{D^2-1\over D^2}a^2+b^2\leq 1\;,
\end{equation}
which means that $\xi_2$ lies on or within an ellipse that has
principal radius $D/\sqrt{D^2-1}\ge1$ along the real axis and
principal radius 1 along the imaginary axis.  Therefore, we
conclude that
\begin{equation}
\label{eq:range1}
0\leq|\xi_2|^2\leq{D^2\over D^2-1}\;.
\end{equation}
It is easy to see that the minimum value of $\xi_1$ occurs when
$\xi_2=D/\sqrt{D^2-1}$, this minimum value being
$\xi_1=-1/\sqrt{D^2-1}$.  It is also easy to see that the maximum
value of $\xi_1$ occurs when $\xi_2$ is real; the maximum occurs
at $\xi_2=-1/\sqrt{D^2-1}$ and is given by $\xi_1=D/\sqrt{D^2-1}$.
The upshot is that $\xi_1$ is bounded by
\begin{equation}
-{1\over\sqrt{D^2-1}}\leq\xi_1\leq{D\over\sqrt{D^2-1}}\;.
\end{equation}
The negative values of $\xi_1$ are unimportant, because they can be
converted to positive values by multiplying both $\xi_1$ and $\xi_2$
by $-1$.  What is important is that $|\xi_1|^2$ has the same range
of possible values as $|\xi_2|^2$.

We now allow qudit $A$ to interact with the two ancilla qudits,
the resulting dynamics described by the unitary operator
\begin{equation}
{U}_{ABC}=\exp[-i(x_C-x_B)p_A]\exp[-ix_A(p_B+p_C)]
\label{uabc}
\end{equation}
(for more details, see Ref.~\cite{Braunstein2000}).  For an
initial pure state $|\psi\rangle$ of qudit $A$, the joint state
after the interaction is
\begin{equation}
{U}_{ABC}|\psi\rangle_A\otimes|\Phi\rangle_{BC} =
\xi_1|\psi\rangle_A\otimes|\Xi_{00}\rangle_{BC}+\xi_2
|\psi\rangle_B\otimes|\Xi_{00}\rangle_{AC}\;.
\end{equation}
The output states of the individual qudits after tracing out the
other two qudits are
\begin{eqnarray}
\label{eq:out1}
{\rho}_A^{(\rm out)} &=&
\left( \xi_1^2+{\xi_1(\xi_2+\xi_2^*)\over D}\right)\!{\rho}
+ {|\xi_2|^2\over D} {I} \;, \\
{\rho}_B^{(\rm out)} &=&
\left( |\xi_2|^2+{\xi_1(\xi_2+\xi_2^*)\over D}\right)\!{\rho}
+{\xi_1^2\over D} {I} \;,  \\
{\rho}_C^{(\rm out)} &=& {\xi_1(\xi_2+\xi_2^*)\over D}{\rho}^T
+{\xi_1^2+|\xi_2|^2\over D}I \;,
\end{eqnarray}
where $\rho$ is an arbitrary initial state of qudit $A$ and
${\rho}^T$ is its transpose.  Taking into account the
constraint~(\ref{eq:cons}), we can rewrite the output states of
qudits $A$ and $B$ as
\begin{eqnarray}
\label{eq:outA}
{\rho}_A^{(\rm out)}
&=&(1-|\xi_2|^2){\rho}+|\xi_2|^2 I/D\;, \\
{\rho}_B^{(\rm out)}&=&(1-\xi_1^2){\rho}+\xi_1^2 I/D \;.
\label{eq:outB}
\end{eqnarray}
As far as qudit $A$ is concerned, the QID acts like the
superoperator ${\cal G}_{DT}$ of Eqs.~(\ref{eq:Gtp}) and
(\ref{eq:Gtprho}) with $D\nu_D=|\xi_2|^2$.  As far as qudit $B$ is
concerned, the QID first swaps the states of $A$ and $B$ and then
acts like ${\cal G}_{DT}$ with $D\nu_D=\xi_1^2$.

Rewriting the output state of qudit $A$ in terms of the ideal inverted
state $\rho^\perp=(I-\rho)/(D-1)$, we get
\begin{equation}
{\rho}_A^{(\rm out)}=
(|\xi_2|^2-1)(D-1)\rho^\perp+[D-|\xi_2|^2(D-1)]I/D\;.
\label{out3}
\end{equation}
To make ${\rho}_A^{(\rm out)}$ as close as possible to $\rho^\perp$,
we need to maximize $|\xi_2|^2$; i.e., we need to choose
\begin{equation}
D\nu_D=|\xi_2|^2={D^2\over D^2-1}\;,
\end{equation}
thus making the action of the QID on qudit $A$ the same as the
action of the universal-NOT given in Eq.~(\ref{eq:inver}).  Notice
that the QID gives the superoperator ${\cal G}_{\rm AV}$ of
Eq.~(\ref{eq:GAV}) when $D\nu_D=|\xi_2|^2=D/(D+1)$.

When $|\xi_2|^2$ has its maximum value, $\xi_1^2=1/(D^2-1)$, so
the output state~(\ref{eq:outB}) of qudit $B$ becomes
\begin{equation}
{\rho}_B^{(\rm out)}=
\left(1-\frac{1}{D^2-1}\right){\rho}+\frac{1}{(D^2-1)}{I\over D}\,.
\label{2.22}
\end{equation}
Notice that in the limit of large $D$, we have
$|\xi_2|\rightarrow1$ and $\xi_1\rightarrow 0$. The output state
of qudit $B$ reduces to the input state of qudit $A$, and the
output states of $A$ and $C$ reduce to the maximally mixed state
${I}/D$.  All this is a consequence of the fact that the initial
state of qudits $B$ and $C$ limits to
$|\Phi\rangle_{BC}\rightarrow|x_0\rangle_B\otimes|p_0\rangle_C$,
and the QID swaps the states of $A$ and $B$:
\begin{equation}
{U}_{ABC}|\psi\rangle_A\otimes|\Xi_{00}\rangle_{BC} =
|\psi\rangle_B\otimes|\Xi_{00}\rangle_{AC}\;.
\end{equation}

Our second realization of the universal-NOT starts with a measurement
of the isotropic POVM
\begin{equation}
dE(|\psi\rangle)=D{d{\cal V}\over{\cal V}}\,|\psi\rangle\langle\psi|\;,
\end{equation}
where
\begin{equation}
\int dE(|\psi\rangle)=
D\int{d{\cal V}\over{\cal V}}\,|\psi\rangle\langle\psi|=I\;.
\end{equation}
We assume that the measurement projects the system onto the measured
state, so the operation that describes a measurement whose result is
the state $|\psi\rangle$ is
\begin{equation}
d{\cal A}(|\psi\rangle)=D{d{\cal V}\over{\cal V}}\,
|\psi\rangle\langle\psi|\odot|\psi\rangle\langle\psi|\;.
\end{equation}
Knowing that the system is in the state $|\psi\rangle$, we can invert
the state.  The operation that describes the measurent followed by
inversion is ${\cal S}_{DT}\circ d{\cal A}(|\psi\rangle)$, where
${\cal S}_{DT}$ is the trace-preserving version of the universal
inverter.  If we now throw away the result of the measurement of the
isotropic POVM, the resulting trace-preserving operation is
\begin{equation}
\int{\cal S}_{DT}\circ d{\cal A}(|\psi\rangle)=
{\cal S}_{DT}\circ{\cal G}_{\rm AV}\;,
\end{equation}
where ${\cal G}_{\rm AV}$ is the superoperator that describes
projection onto a random pure state [see Eq.~(\ref{eq:GAV2})].

Using the forms~(\ref{eq:tpuinverter}) and (\ref{eq:GAV}), we can
write the overall operation as
\begin{equation}
{\cal S}_{DT}\circ{\cal G}_{\rm AV}=
{1\over D^2-1}({\bf I}-{\cal I})\circ({\bf I}+{\cal I})=
{1\over D^2-1}(D{\bf I}-{\cal I})={\cal G}_{\rm NOT}\;,
\end{equation}
where we use the fact that ${\bf I}\circ{\bf I}=D{\bf I}$.  This
demonstrates that the universal-NOT results from a measurement
of the isotropic POVM followed by state inversion.

\section{Conclusion}
\label{sec:conclusion}

The concurrence introduced by Hill and Wootters \cite{Hill1997} and
by Wootters \cite{Wootters1998} provides a good measure of the
entanglement of any state of two qubits, pure or mixed.  The
Hill-Wootters concurrence is generated with the help of the
superoperator that flips the spin of a qubit.  In this paper we have
identified the crucial properties of the spin-flip superoperator,
which allow it to generate a good entanglement measure for pure
states of two qubits.  By generalizing these properties to systems of
arbitrary dimension, we have singled out a unique superoperator,
which we call the universal inverter.  In the same way that the spin
flip generates a concurrence for pairs of qubits, the universal
inverter generates a concurrence, which we call the I-concurrence,
for joint pure states of pairs of quantum systems of arbitrary
dimension. This pure-state I-concurrence measures entanglement in
terms of the purity of the marginal density operators of the joint
pure state.

It is natural to define the I-concurrence of mixed states of
$D_1\times D_2$ quantum systems as the minimum average I-concurrence
of ensemble decompositions of the joint density operator.
Property~$3'$ of the I-concurrence---that the I-concurrence of a pure
state $|\Psi\rangle$ is zero if and only if $|\Psi\rangle$ is a
product state---implies immediately that the mixed-stated concurrence
just defined is zero if and only if the mixed state is separable.  We
are investigating further properties of this mixed-state
I-concurrence and how it is related to other measures of mixed-state
entanglement.

The universal inverter turns out to be the ideal inverter of pure
states, since it takes a pure state to the maximally mixed state in
the subspace orthogonal to the pure state.  Because the universal
inverter is a positive, but not completely superoperator, it cannot
be realized as the dynamics of a quantum system coupled to an
ancilla.  We have shown that among a one-parameter family of
inverting superoperators, the completely positive superoperator that
comes closest to achieving an ideal state inversion is a
superoperator called the universal-NOT, and we have presented a
physical realization of the universal-NOT.

\acknowledgements
This work was supported in part by the Office of
Naval Research (Grant No.~N00014-00-1-0578), the EQUIP project
of the European Union 5th Framework research program, Information
Society Technologies (Contract No.~IST-1999-11053), and the
National Science Foundation (Grant No.~PHY-9970507).

\appendix

\section{Superoperator formalism and special superoperators}
\label{app:supop}

The formalism we use for superoperators has been used extensively
in open-systems theory \cite{Alicki1987}.  In this Appendix, we
summarize our notation, which follows that of
Ref.~\cite{Caves1999}, and we introduce and describe key
properties of several superoperators that are important for our
analysis.

The space of linear operators acting on a Hilbert space ${\cal H}$
is a $D^2$-dimensional complex vector space.  We introduce
operator ``kets'' $|A)=A$ and ``bras'' $(A|=A^\dagger$,
distinguished from vector kets and bras by the use of smooth
brackets.  The natural operator inner product can be written as
$(A|B)={\rm tr}(A^\dagger B)$.  An orthonormal basis $|e_j\rangle$
induces an orthonormal operator basis
\begin{equation}
|e_j\rangle\langle e_k|=\tau_{jk}\equiv\tau_\alpha\;,
\end{equation}
where the Greek index is an abbreviation for two Roman indices.
Not all orthonormal operator bases are of this outer-product form.
In the following, $\tau_\alpha$ can be a general orthonormal
operator basis, or it can be specialized to an outer-product
basis.

The space of superoperators on ${\cal H}$, i.e., linear maps on
operators, is a $D^4$-dimensional complex vector space.  A
superoperator ${\cal A}$ is specified by its ``matrix elements''
\begin{equation}
{\cal A}_{lj,mk}
\equiv
\Bigl\langle e_l\Bigl|
{\cal A}(|e_j\rangle\langle e_k|)
\Bigr|e_m\Bigr\rangle\;,
\end{equation}
for the superoperator can be written in terms of its matrix
elements as
\begin{equation}
{\cal A}=\sum_{lj,mk}
{\cal A}_{lj,mk}
|e_l\rangle\langle e_j|\odot|e_k\rangle\langle e_m|
=
\sum_{\alpha,\beta}
{\cal A}_{\alpha\beta}\,\tau_\alpha\odot\tau_\beta^\dagger=
\sum_{\alpha,\beta}
{\cal A}_{\alpha\beta}
|\tau_\alpha)(\tau_\beta|\;.
\end{equation}
The {\it ordinary action\/} of ${\cal A}$ on an operator $A$, used
above to generate the matrix elements, is obtained by dropping an
operator $A$ into the center of the representation of ${\cal A}$,
in place of the $\odot$ sign, i.e.,
\begin{equation}
{\cal A}(A)=
\sum_{\alpha,\beta}
{\cal A}_{\alpha\beta}\,\tau_\alpha A\tau_\beta^\dagger\;.
\end{equation}
There is clearly another way that ${\cal A}$ can act on $A$, the
{\it left-right action},
\begin{equation}
{\cal A}|A)\equiv
\sum_{\alpha,\beta}
{\cal A}_{\alpha\beta}
|\tau_\alpha)(\tau_\beta|A)\;,
\end{equation}
in terms of which the matrix elements are
\begin{equation}
{\cal A}_{\alpha\beta}
=(\tau_\alpha|\,{\cal A}|\tau_\beta)
=\Bigl(|e_l\rangle\langle e_j|\Bigl|
{\cal A}\Bigr||e_m\rangle\langle e_k|\Bigr)
=\Bigl\langle e_l\Bigl|
{\cal A}(|e_j\rangle\langle e_k|)
\Bigr|e_m\Bigr\rangle
={\cal A}_{lj,mk}
\;.
\label{eq:fundconn}
\end{equation}
This expression provides the fundamental connection between the two
actions of a superoperator.

With respect to the left-right action, a superoperator works just
like an operator.  Multiplication of superoperators ${\cal B}$ and
${\cal A}$ is given by
\begin{equation}
{\cal B\cal A}=
\sum_{\alpha,\beta,\gamma}
{\cal B}_{\alpha\gamma}{\cal A}_{\gamma\beta}
|\tau_\alpha)(\tau_\beta|\;,
\end{equation}
and the ``left-right'' adjoint, defined by
\begin{equation}
(A|{\cal A}^\dagger|B)=(B|{\cal A}|A)^*\;,
\label{eq:lradjoint}
\end{equation}
is given by
\begin{equation}
{\cal A}^\dagger=
\sum_{\alpha,\beta}
{\cal A}_{\alpha\beta}^*
\tau_\beta\odot\tau_\alpha^\dagger=
\sum_{\alpha,\beta}
{\cal A}_{\beta\alpha}^*
|\tau_\alpha)(\tau_\beta|\;.
\end{equation}
With respect to the ordinary action, superoperator multiplication,
denoted as a composition ${\cal B}\circ{\cal A}$, is given by
\begin{equation}
{\cal B}\circ{\cal A}=
\sum_{\alpha,\beta,\gamma,\delta}
{\cal B}_{\gamma\delta}{\cal A}_{\alpha\beta}\,
\tau_\gamma\tau_\alpha
\odot\tau_\beta^\dagger\tau_\delta^\dagger\;.
\end{equation}
The adjoint with respect to the ordinary action, denoted by
${\cal A}^\times$, is defined by
\begin{equation}
{\rm tr}\Bigl([{\cal A}^\times(B)]^\dagger A\Bigr)=
{\rm tr}\Bigl(B^\dagger{\cal A}(A)\Bigr)\;.
\label{eq:ordadjoint}
\end{equation}
In terms of a representation in an operator basis, this
``cross'' adjoint becomes
\begin{equation}
{\cal A}^\times=
\sum_{\alpha,\beta}
{\cal A}_{\alpha\beta}^*\,
\tau_\alpha^\dagger\odot\tau_\beta\;.
\end{equation}
Notice that
\begin{equation}
({\cal B}\circ{\cal A})^\dagger={\cal B}^\dagger\circ{\cal A}^\dagger
\quad{\rm and}\quad
({\cal B}{\cal A})^\times={\cal B}^\times{\cal A}^\times\;.
\end{equation}

We can formalize the connection between the two kinds of action by
defining an operation, called ``sharp,'' which exchanges the two:
\begin{equation}
{\cal A}^\#|A)\equiv{\cal A}(A)\;.
\end{equation}
Simple consequences of the definition are that
\begin{eqnarray}
&\mbox{}&({\cal A}^\#)^\dagger=({\cal A}^\times)^\#\;,\\
&\mbox{}&({\cal B}\circ{\cal A})^\#={\cal B}^\#{\cal A}^\#\;.
\label{eq:sharpmult}
\end{eqnarray}
The matrix elements of ${\cal A}^\#$ are given by
\begin{eqnarray}
{\cal A}^\#_{lj,mk}&=&
\Bigl(|e_l\rangle\langle e_j|\Bigl|
{\cal A}^\#\Bigr||e_m\rangle\langle e_k|\Bigr) \nonumber\\
&=&{\rm tr}\Bigl(|e_j\rangle\langle e_l|
{\cal A}(|e_m\rangle\langle e_k|)\Bigr) \nonumber\\
&=&\Bigl\langle e_l\Bigl|
{\cal A}(|e_m\rangle\langle e_k|)
\Bigr|e_j\Bigr\rangle \nonumber\\
&=&{\cal A}_{lm,jk}\;,
\end{eqnarray}
which implies that
\begin{equation}
{\cal A}^\#=
\sum_{lj,mk}
{\cal A}_{lj,mk}
|e_l\rangle\langle e_m|\odot|e_k\rangle\langle e_j|\;.
\end{equation}

A superoperator is left-right Hermitian, i.e., ${\cal A}^\dagger={\cal A}$,
if and only if it has an eigendecomposition
\begin{equation}
{\cal A}=
\sum_{\alpha}\mu_\alpha|\tau_\alpha)(\tau_\alpha|
=\sum_{\alpha}\mu_\alpha\tau_\alpha\odot\tau_\alpha^\dagger\;,
\label{eq:Aeigendecomp}
\end{equation}
where the $\mu_\alpha$ are real (left-right) eigenvalues and the
operators $\tau_\alpha$ are orthonormal eigenoperators.

A superoperator is {\it trace preserving\/} if, under the ordinary
action, it leaves the trace unchanged, i.e., if ${\rm tr}(A)= {\rm
tr}({\cal A}(A))={\rm tr}([{\cal A}^\times(I)]^\dagger A)$ for all
operators $A$.  Thus ${\cal A}$ is trace preserving if and only if
${\cal A}^\times(I)=I$.

A superoperator is said to be {\it positive\/} if it maps positive
operators to positive operators under the ordinary action.  A
superoperator is {\it completely positive\/} if it and all its
extensions ${\cal I}\otimes{\cal A}$ to tensor-product spaces,
where ${\cal I}$ is the unit superoperator on the appended space,
are positive.  It can be shown that ${\cal A}$ is completely
positive if and only if it is positive relative to the left-right
action, i.e., $(A|{\cal A}|A)\ge0$ for all operators $A$ (for a
proof in the present notation, see Ref.~\cite{Caves1999}).  This
is equivalent to saying that ${\cal A}$ is left-right Hermitian
with nonnegative left-right eigenvalues.

In this paper we make use of several special superoperators, whose
properties we summarize here.  The identity superoperator with
respect to the ordinary action is
\begin{equation}
{\cal I}=I\odot I=
\sum_{j,k}|e_j\rangle\langle e_j|\odot|e_k\rangle\langle e_k|\;.
\end{equation}
This superoperator is Hermitian in both senses, i.e., ${\cal
I}={\cal I}^\dagger={\cal I}^\times$.  It is the identity
superoperator relative to the ordinary action because ${\cal
I}(A)=A$ for all operators $A$, but its left-right action gives
${\cal I}|A)={\rm tr}(A)I$.

The identity superoperator with respect to the left-right action
is
\begin{equation}
{\bf I}=\sum_\alpha|\tau_\alpha)(\tau_\alpha|
=\sum_{j,k}|e_j\rangle\langle e_k|\odot|e_k\rangle\langle e_j|\;.
\end{equation}
This superoperator is also Hermitian in both senses, i.e., ${\bf
I}={\bf I}^\dagger={\bf I}^\times$.  It is the identity
superoperator relative to the left-right action because ${\bf
I}|A)=A$ for all operators $A$, but its ordinary action gives
${\bf I}(A)={\rm tr}(A)I$.  Since sharping exchanges the two
kinds of action, it is clear that ${\cal I}^\#={\bf I}$.

To define the remaining superoperators, it is useful to introduce
a set of $D^2-1$ tracefree, Hermitian operators \cite{Lendi1987},
which are the generators of SU($D$).  We label these operators by
a Greek index $\alpha$, which runs from 1 to $D^2-1$.  The
operators are defined by
\begin{eqnarray}
&\mbox{}&
\alpha=1,\ldots,D-1:\nonumber\\
&\mbox{}&\hphantom{\alpha=1,\ldots,}
   \lambda_\alpha=\Gamma_j\equiv{1\over\sqrt{j(j-1)}}
   \left(\,\sum_{k=1}^{j-1}\tau_{kk}-(j-1)\tau_{jj}\right)
   \;,\quad 2\leq j\leq D\;,\label{eq:diagonal}\\
&\mbox{}&
\alpha=D,\ldots,{(D+2)(D-1)/2}:\nonumber\\
&\mbox{}&\hphantom{\alpha=1,\ldots,}
    \lambda_\alpha=\Gamma_{jk}^{(+)}\equiv{1\over\sqrt2}
    \left(\tau_{jk}+\tau_{kj}\right)
    \;,\quad 1\leq j<k\leq D\;,\label{eq:plus}\\
&\mbox{}&
\alpha=D(D+1)/2,\ldots,D^2-1:\nonumber\\
&\mbox{}&\hphantom{\alpha=1,\ldots,}
    \lambda_\alpha=\Gamma_{jk}^{(-)}\equiv{-i\over\sqrt2}
    \left(\tau_{jk}-\tau_{kj}\right)
    \;,\quad 1\leq j<k\leq D\;.\label{eq:minus}
\end{eqnarray}
In Eq.~(\ref{eq:diagonal}), $\alpha$ stands for a single Roman
index $j$, whereas in Eqs.~(\ref{eq:plus}) and (\ref{eq:minus}),
it stands for the pair of Roman indices, $jk$. These operators are
Hermitian generalizations of the two-dimensional Pauli operators:
the operators~(\ref{eq:diagonal}) are diagonal in the chosen
basis, like $\sigma_z$; for each pair of dimensions, the
operators~(\ref{eq:plus}) are like the Pauli operator $\sigma_x$;
and for each pair of dimensions, the operators~(\ref{eq:minus})
are like $\sigma_y$.

Like the Pauli operators, the operators $\lambda_\alpha$ are
orthonormal, i.e.,
\begin{equation}
(\lambda_\alpha|\lambda_\beta)=
{\rm tr}(\lambda_\alpha\lambda_\beta)=\delta_{\alpha\beta}\;.
\end{equation}
Thus they constitute an operator basis for the subspace of
tracefree operators.  Indeed, we can define a superoperator
projector,
\begin{equation}
{\cal F}\equiv\sum_\alpha|\lambda_\alpha)(\lambda_\alpha|=
\sum_\alpha\lambda_\alpha\odot\lambda_\alpha\;,
\end{equation}
which relative to the left-right action, projects onto the
subspace of tracefree operators.  Notice that ${\cal F}= {\cal
F}^\dagger={\cal F}^\times$.

If we add to the set of operators $\lambda_\alpha$ the normalized
unit operator $I/\sqrt D$, we obtain an orthonormal operator basis.
Thus the unit superoperator ${\bf I}$ can be written as
\begin{equation}
{\bf I}={|I)(I|\over D}+\sum_\alpha|\lambda_\alpha)(\lambda_\alpha|
={\cal I}/D+{\cal F}\;.
\end{equation}
Writing ${\cal F}={\bf I}-{\cal I}/D$, we find that
\begin{equation}
{\cal F}^\#={\cal I}-{{\bf I}\over D}
={D^2-1\over D^2}{\cal I}-{\cal F\over D}\;.
\end{equation}

In the chosen basis, the operators~(\ref{eq:diagonal}) and
(\ref{eq:plus}) are real and symmetric.  Together with $I/\sqrt
D$, they constitute a set of $D(D+1)/2$ orthonormal operators,
which span the subspace of operators that are symmetric in the
chosen basis.  In contrast, the $D(D-1)/2$ operators in
Eq.~(\ref{eq:minus}) are pure imaginary and antisymmetric and span
the subspace of operators that are antisymmetric in the chosen
basis.  We can define superoperator projectors,
\begin{eqnarray}
{\cal P}_S&\equiv&
{|I)(I|\over D}+
\sum_{\lambda_\alpha\;\mbox{\scriptsize real}}
|\lambda_\alpha)(\lambda_\alpha|
\;,\\
{\cal P}_A&\equiv&
\sum_{\lambda_\alpha\;\mbox{\scriptsize imaginary}}
|\lambda_\alpha)(\lambda_\alpha|\;,
\end{eqnarray}
which relative to the left-right action, project onto the
symmetric and antisymmetric operator subspaces. Notice that ${\cal
P}_S={\cal P}_S^\dagger={\cal P}_S^\times$ and ${\cal P}_A={\cal
P}_A^\dagger={\cal P}_A^\times$.  It is clear that
\begin{equation}
{\bf I}={\cal P}_S+{\cal P}_A\;.
\label{eq:ISA}
\end{equation}

The last superoperator we need is the superoperator that
transposes operators in the chosen basis.  The ordinary action of
the transposition superoperator is given by
\begin{equation}
{\cal T}(A)=\sum_{j,k}|e_j\rangle\langle e_k|A|e_j\rangle\langle e_k|\;,
\end{equation}
so the superoperator has the form
\begin{equation}
{\cal T}=\sum_{j,k}|e_j\rangle\langle e_k|\odot|e_j\rangle\langle e_k|\;.
\end{equation}
The transposition superoperator is Hermitian in both senses and is
unchanged by sharping, i.e., ${\cal T}={\cal T}^\dagger={\cal
T}^\times={\cal T}^\#$.  In addition to satisfying ${\cal
T}\circ{\cal T}={\cal I}$, the transposition superoperator has the
property that
\begin{equation}
{\bf I}\circ{\cal T}={\bf I}\;,
\label{eq:IT}
\end{equation}
which in view of Eq.~(\ref{eq:sharpmult}), is equivalent to
${\cal IT}={\cal I}$.

It is easy to see that ${\cal P}_S-{\cal P}_A$, acting to the right,
transposes an operator, i.e.,
\begin{equation}
{\cal P}_S|A)-{\cal P}_A|A)={\cal T}(A)={\cal T}^\#|A)\;,
\end{equation}
which gives us, since ${\cal T}$ is invariant under sharping,
\begin{equation}
{\cal T}={\cal T}^\#={\cal P}_S-{\cal P}_A\;.
\end{equation}
Combined with Eq.~(\ref{eq:ISA}), this gives us
\begin{eqnarray}
{\cal P}_S&=&{1\over2}({\bf I}+{\cal T})\;,\\
{\cal P}_A&=&{1\over2}({\bf I}-{\cal T})\;.
\end{eqnarray}
Combining these forms with Eq.~(\ref{eq:IT}) yields
\begin{eqnarray}
2{\cal P}_S\circ{\cal T}&=&{\bf I}+{\cal I}=(D+1){\cal G}_{\rm AV}\;,\\
2{\cal P}_A\circ{\cal T}&=&{\bf I}-{\cal I}={\cal S}_D/\nu_D\;.
\label{eq:PAT}
\end{eqnarray}
The form~(\ref{eq:PAT}) has been given previously \cite{Horodecki1999}.

\section{}
\label{app:HtoH}

In this Appendix we show that a superoperator is Hermitian
relative to the left-right action if and only if it maps all
Hermitian operators to Hermitian operators.

Let $\A$ be a superoperator, and let $|e_j\rangle$ be an
orthonormal basis, which induces an orthonormal operator basis
$|e_j\rangle\langle e_k|$. Notice that
\begin{eqnarray}
\Bigl\langle e_l\Bigl|
\A^\dagger(|e_j\rangle\langle e_k|)
\Bigr|e_m\Bigr\rangle
& = &
\Bigl(\,|e_l\rangle\langle e_j|\,\Bigl|
\A^\dagger\Bigr|\,|e_m\rangle\langle e_k|\,\Bigr)\nonumber\\
& = &
\Bigl(\,|e_m\rangle\langle e_k|\,\Bigl|
\A\Bigr|\,|e_l\rangle\langle e_j|\,\Bigr)^*\nonumber\\
& = &
\Bigl\langle e_m\Bigl|
\A(|e_k\rangle\langle e_j|)
\Bigr|e_l\Bigr\rangle^*\nonumber\\
& = &
\Bigl\langle e_l\Bigl|
[\A(|e_k\rangle\langle e_j|\,)]^\dagger
\Bigr|e_m\Bigr\rangle\;.
\label{eq:adjrel}
\end{eqnarray}
Here the first and third equalities follow from relating the
ordinary action of a superoperator to its left-right action
[Eq.~(\ref{eq:fundconn})], the second equality follows from the
definition of the left-right adjoint of $\A$
[Eq.~(\ref{eq:lradjoint})], and the fourth equality follows from
the definition of the operator adjoint. Equation~(\ref{eq:adjrel})
gives the relation between the operator adjoint and the left-right
superoperator adjoint:
\begin{equation}
\A^\dagger(|e_j\rangle\langle e_k|)
=[\A(|e_k\rangle\langle e_j|)]^\dagger\;.
\end{equation}
Thus we have that $\A=\A^\dagger$, i.e., $\A$ is left-right
Hermitian, if and only if
\begin{equation}
\A(|e_j\rangle\langle e_k|)
=[\A(|e_k\rangle\langle e_j|)]^\dagger
\label{eq:Hermrel}
\end{equation}
for all $j$ and $k$. This result allows us to prove the desired
theorem easily.

{{\bf Theorem.} A superoperator $\A$ is left-right Hermitian if
and only if it maps all Hermitian operators to Hermitian
operators.

{\sl Proof:}  First suppose $\A$ is left-right Hermitian, i.e.,
$\A=\A^\dagger$.  This implies that $\A$ has a complete,
orthonormal set of eigenoperators $\tau_\alpha$, with real
eigenvalues $\mu_\alpha$.  Using the
eigendecomposition~(\ref{eq:Aeigendecomp}), we have for any
Hermitian operator $H$,
\begin{equation}
\A(H)
=\sum_\alpha\mu_\alpha\tau_\alpha H\tau_\alpha^\dagger
=\A(H)^\dagger\;.
\end{equation}

Now suppose $\A$ maps all Hermitian operators to Hermitian
operators.  Letting $\tau_{jk}=|e_j\rangle\langle e_k|$, it
follows that
\begin{eqnarray}
\A(\tau_{jk})
& = &\A\!\left(
{1\over2}(\tau_{jk}+\tau_{kj})+i{-i\over2}(\tau_{jk}-\tau_{kj})
\right)\nonumber\\
& = & \A\!\left(
{1\over2}(\tau_{jk}+\tau_{kj})
\right)
+i\A\!\left(
{-i\over2}(\tau_{jk}-\tau_{kj})
\right)\nonumber\\
& = & \left[\A\!\left(
{1\over2}(\tau_{jk}+\tau_{kj})
\right)\right]^\dagger
+i\!\left[\A\!\left(
{-i\over2}(\tau_{jk}-\tau_{kj})
\right)\right]^\dagger\nonumber\\
& = & \left[\A\!\left(
{1\over2}(\tau_{jk}+\tau_{kj})
\right)
-i\A\!\left(
{-i\over2}(\tau_{jk}-\tau_{kj})
\right)\right]^\dagger\nonumber\\
& = & \left[\A\!\left(
{1\over2}(\tau_{jk}+\tau_{kj})-i{-i\over2}(\tau_{jk}-\tau_{kj})
\right)\right]^\dagger\nonumber\\
& = & [\A(\tau_{kj})]^\dagger\;.
\end{eqnarray}
Equation~(\ref{eq:Hermrel}) then implies that $\A=\A^\dagger$.$\QED$

Since a superoperator is left-right Hermitian if and only if it
has an eigendecomposition as in Eq.~(\ref{eq:Aeigendecomp}), we
can conclude, by grouping together positive and negative
eigenvalues, that being left-right Hermitian is equivalent to
being the difference between two completely positive
superoperators.  Using the theorem, we have that a superoperator
takes all Hermitian operators to Hermitian operators if and only
if it is the difference between two completely positive
superoperators.  This generalizes a result of Yu \cite{Yu2000},
who showed that a positive superoperator is the difference between
two completely positive superoperators.  From our perspective,
we can say that since a positive superoperator takes positive
operators to positive operators, it also takes Hermitian operators
to Hermitian operators and thus is left-right Hermitian.  A
positive operator that is not completely positive has one or
more negative left-right eigenvalues.

We can get one further result relevant to the considerations in
this paper: if ${\cal A}$ and ${\cal B}$ are left-right Hermitian
superoperators for two separate quantum systems, then ${\cal
A}\otimes{\cal B}$ is also left-right Hermitian and thus maps all
Hermitian operators of the joint system to Hermitian operators.

\section{}
\label{app:rho}

Let
\begin{equation}
\rho_A=\sum_{j=1}^{D_1}\mu_j|e_j\rangle\langle e_j|
\quad\hbox{and}\quad
\rho_B=\sum_{k=1}^{D_2}\nu_k|f_k\rangle\langle f_k|
\end{equation}
be the eigendecompositions of $\rho_A$ and $\rho_B$.  In the joint basis
$|e_j\rangle\otimes|f_k\rangle$, $\rho_{AB}$ has the form
\begin{equation}
\rho_{AB}=
\sum_{j,k,l,m}\rho_{jk,lm}
|e_j\rangle\langle e_l|\otimes|f_k\rangle\langle f_m|\;.
\end{equation}
The diagonal forms of the marginal density operators show that
\begin{equation}
\sum_{k=1}^{D_2}\rho_{jk,lk}=\mu_j\delta_{jl}
\quad\hbox{and}\quad
\sum_{j=1}^{D_1}\rho_{jk,jm}=\nu_k\delta_{km}\;.
\end{equation}
Thus the diagonal elements of $\rho_{jk,lm}$ are a probability
distribution $p_{jk}=\rho_{jk,jk}$, whose marginals are the
eigenvalues of the marginal density operators:
\begin{equation}
\sum_{k=1}^{D_2} p_{jk}=\mu_j
\quad\hbox{and}
\quad\sum_{j=1}^{D_1} p_{jk}=\nu_k\;.
\end{equation}
We now can write
\begin{eqnarray}
1+{\rm tr}(\rho_{AB}^2)
& = & 1+\sum_{j,k,l,m}|\rho_{jk,lm}|^2\cr \nonumber \\
& \ge  & 1+\sum_{j,k}p_{jk}^2\cr \nonumber \\
& = & \sum_{j,k,l,m}p_{jk}p_{lm}+\sum_{j,k}p_{jk}^2\cr \nonumber \\
& = & \sum_{j,k,m}p_{jk}p_{jm}+\sum_{j\ne l,k,m}p_{jk}p_{lm}
    +\sum_{j,k,l}p_{jk}p_{lk}-\sum_{j\ne l,k}p_{jk}p_{lk}\cr \nonumber \\
& = & \sum_j\biggl(\sum_k p_{jk}\biggr)^2+
    \sum_k\biggl(\sum_j p_{jk}\biggr)^2+
    \sum_{j\ne l,k\ne m}p_{jk}p_{lm} \nonumber \\
& \ge & \sum_j\mu_j^2+\sum_l\nu_k^2 \nonumber \\
& = & {\rm tr}(\rho_A^2)+{\rm tr}(\rho_B^2)\;.
\label{eq:rhoineq}
\end{eqnarray}

The first inequality here is saturated if and only if $\rho_{AB}$
is diagonal in the  basis $|e_j\rangle\otimes|f_k\rangle$.   The
second inequality is saturated if and only if $p_{jk}p_{lm}=0$
whenever $j\ne l$ and $k\ne m$}.  This requirement is equivalent
to saying that the nonzero entries in $p_{jk}$ are restricted to
one row or to one column.  In view of the first requirement, this
means that overall equality is achieved in Eq.~(\ref{eq:rhoineq})
if and only if $\rho_{AB}=\rho_A\otimes\rho_B$ is a product state,
with $\rho_A$ or $\rho_B$ a pure state.

\section{}
\label{app:eigensubspaces}

In this Appendix we show that the vector space of operators acting
on a $D$-dimensional Hilbert space has only two proper operator
subspaces that are invariant under all unitary transformations.
These two subspaces are the one-dimensional subspace spanned by
the unit operator $I$ and the subspace consisting of all tracefree
operators.

It is obvious that the subspace consisting of multiples of $I$ and
the subspace of trace-free operators are unitarily invariant.  To
show that these are the only unitarily invariant proper subspaces,
we consider a unitarily invariant subspace that is not the
subspace spanned by $I$, and we show that this subspace is either
the subspace of tracefree operators or the entire operator space.
Let $A$ be a nonzero operator in the unitarily invariant subspace,
which is not a multiple of $I$.  There exists an orthonormal basis
$|e_j\rangle$ such that $A_{11}\ne A_{22}$. Adopt this basis, in
which $A$ has the representation
\begin{equation}
A=\sum_{j,k=1}^D A_{jk}|e_j\rangle\langle e_k|\;.
\end{equation}

Consider the unitary operator $U$ that changes the sign of
$|e_1\rangle$, i.e., $U|e_1\rangle=-|e_1\rangle$ and
$U|e_j\rangle=|e_j\rangle$ for $j=2,\ldots,D$.  Also in the
unitarily invariant subspace is the operator
\begin{equation}
B={1\over2}(A+UAU^\dagger)=A_{11}|e_1\rangle\langle e_1|+
\sum_{j,k=2}^D A_{jk}|e_j\rangle\langle e_k|\;.
\end{equation}
Do the same thing to the second basis vector; i.e., use the
unitary operator $V$ defined by $V|e_2\rangle=-|e_2\rangle$, and
$V|e_j\rangle=|e_j\rangle$ for $j=1$ and $j=3,\ldots,D$.  Also in
the subspace is the operator
\begin{equation}
C={1\over2}(B+VBV^\dagger)=A_{11}|e_1\rangle\langle e_1|
+A_{22}|e_2\rangle\langle e_2|+
\sum_{j,k=3}^D A_{jk}|e_j\rangle\langle e_k|\;.
\end{equation}
Now consider the unitary operator $W$ that swaps $|e_1\rangle$ and
$|e_2\rangle$, i.e., $W|e_1\rangle=|e_2\rangle$,
$W|e_2\rangle=|e_1\rangle$, and $W|e_j\rangle=|e_j\rangle$ for
$j=3,\ldots,D$.  Also in the subspace is the (nonzero) tracefree
operator
\begin{equation}
D=C-WCW^\dagger=(A_{11}-A_{22})
(|e_1\rangle\langle e_1|-|e_2\rangle\langle e_2|)\;.
\end{equation}

We conclude that the subspace contains the tracefree operator
$|e_1\rangle\langle e_1|-|e_2\rangle\langle e_2|$, which is a Pauli
$\sigma_z$ operator for the first two dimensions.  From this
operator, we can generate by unitary transformations that
interchange basis vectors a $\sigma_z$-like operator for every
pair of dimensions, and from these $\sigma_z$ operators, we can
generate by unitary transformations a $\sigma_x$ and a $\sigma_y$
operator for every pair of dimensions.  Since these Pauli-like
operators span the space of tracefree operators, we conclude that
any unitarily invariant operator subspace that is not the space
spanned by $I$ contains all tracefree operators.

The unitarily invariant subspace could be the subspace of
tracefree operators.  Suppose that it is not and thus contains an
operator $E$ that is not tracefree.  Defining a tracefree operator
$F=E-{\rm tr}(E)I/D$, we see that $I$ can be written as linear
combination of $F$ and $E$ and thus is in the subspace.  Since the
tracefree operators together with $I$ span the entire space of
operators, we conclude that in this case the unitarily invariant
subspace is the entire operator space. This establishes our
result.


\begin{thebibliography}{99}

\bibitem[*]{padd}Permanent address:
Institute of Physics, Slovak Academy of Sciences,
D\'ubravsk\'a cesta 9, 842 28 Bratislava, Slovakia, and
Faculty of Informatics, Masaryk University, Botanick\'a 68a,
602 00 Brno, Czech Republic

\bibitem{Lo1998}
{\it Introduction to Quantum Computation and Information}, edited
by H.-K.~Lo, S.~Popescu, and T.~Spiller (World Scientific,
Singapore, 1998).

\bibitem{Nielsen2000}
M.~A. Nielsen and I.~L.~Chuang, {\it Quantum Computation and Quantum
Information} (Cambridge University Press, Cambridge, England, 2000).

\bibitem{Bennett1996}
C.~H. Bennett, H.~J. Bernstein, S.~Popescu, and B.~Schumacher,
Phys.\ Rev.~A\, {\bf 53}, 2046 (1996).

\bibitem{Cbennett1996}
C.~H. Bennett, D.~P. DiVincenzo, J.~A. Smolin, and W.~K. Wootters,
Phys.\ Rev.~A {\bf 54}, 3824 (1996).

\bibitem{Hill1997}
S.~Hill and W.~K. Wootters, Phys.\ Rev.\ Lett.\ {\bf 78}, 5022
(1997).

\bibitem{Wootters1998}
W.~K. Wootters, Phys.\ Rev.\ Lett. {\bf 80}, 2245 (1998).

\bibitem{Uhlmann2000}
A.~Uhlmann, Phys.\ Rev.~A {\bf 62}, 032307 (2000).

\bibitem{Horodecki1999}
M.~Horodecki and P.~Horodecki, Phys.\ Rev.~A {\bf 59}, 4206 (1999).

\bibitem{Buzek1999}
V.~Bu\v{z}ek, M.~Hillery, and R.~F.\ Werner, Phys.\ Rev.\ A {\bf 60},
R2626 (1999).

\bibitem{Buzek2000}
V.~Bu\v{z}ek, M.~Hillery, and  R.~F.\ Werner, J.\ Mod.\ Opt.\ {\bf
47}, 211 (2000).

\bibitem{Braunstein2000}
S. L. Braunstein, V. Bu\v{z}ek, and M. Hillery, Phys.\ Rev.~A {\bf 63}, 052313 (2001).

\bibitem{Alicki1987}
R.~Alicki and K.~Lendi, {\it Quantum Dynamical Semigroups and
Applications\/} (Springer, Berlin, 1987).

\bibitem{Caves1999}
C.~M. Caves, J.\ Superconductivity {\bf 12}, 707 (1999).

\bibitem{Rungta2000}
P.~Rungta, W.~J. Munro, K.~Nemoto, P.~Deuar, G.~J. Milburn, C.~M. Caves,
In Directions in Quantum Optics: A Collection of
Papers Dedicated to the Memory of Dan Walls, edited by H.~J.
Carmichael, R.~J. Glauber, M.~O. Scully (Springer-Verlag, Berlin,
2000), pages~149--164.

\bibitem{Messiah1968}
A.~Messiah, {\it Quantum Mechanics, Vol.~II} (North-Holland, Amsterdam,
1968), Chap.~XV.

\bibitem{Derka98}
R.~Derka, V.~Bu\v{z}ek, and  A.~Ekert, Phys.\  Rev.\  Lett.\ {\bf
80}, 1571 (1998).

\bibitem{Lendi1987}
K.~Lendi, J.\ Phys.~A {\bf 20}, 15 (1987).

\bibitem{Yu2000}
S.~Yu, unpublished, {\tt e-print quant-ph/0001053}.

\end{thebibliography}
\end{document}